\documentclass[final,5p,times,twocolumn,authoryear]{elsarticle}

\usepackage{amssymb}
\usepackage{lipsum}
\usepackage[colorlinks=true, linkcolor=blue, citecolor=blue, urlcolor=blue]{hyperref}

\usepackage{graphicx} 
\usepackage{amsmath} 
\usepackage{subcaption}
\usepackage{placeins}
\usepackage[table]{xcolor}
\usepackage[nameinlink,noabbrev]{cleveref}
\creflabelformat{equation}{#2\textup{(#1)}#3}
\crefname{figure}{Fig.}{Figs.}
\crefname{table}{Table}{Tables}

\journal{High Energy Astrophysics}

\begin{document}

\begin{frontmatter}





\title{Imploding Remnants: detection bias against AGNs in massive clusters}

\author[first]{Ross J. Turner}
\author[second]{Georgia S. C. Stewart}
\affiliation[first,second]{organization={School of Natural Sciences, University of Tasmania},
            addressline={Private Bag 37}, 
            city={Hobart},
            postcode={7001}, 
            country={Australia}}
            




\begin{abstract}
We propose that an observed scarcity of remnant lobed AGNs in dense clusters results from a peculiarity in their dynamics upon the cessation of jet activity: a rapid `implosion' of lobes that, in their active phase, were primarily supported by the momentum flux of the jet. We investigate this behaviour by analysing the asymptotic behaviour of the RAiSE dynamical model and comparing our predictions both to the full model and hydrodynamic simulations. We find that remnant lobes powered by weak jets in massive clusters are unstable to implosion on the order of at most a few Myr. Consequently, remnant AGNs in massive clusters ($M_\text{halo} \sim 10^{14.5}$~M$_\odot$) will be under-counted by a factor of at least five compared to those in poorer groups ($M_\text{halo} \sim 10^{12}$~M$_\odot$). The lack of such remnants in observed populations may lead to a significant underestimate of the AGN feedback provided by low-powered jets, especially given their prevalence towards cluster cores where feedback is most effective. We discuss the influence of a stabilising magnetic field sheath on the nature of the implosion: does the lobe cleanly implode in on itself, or do fluid instabilities turbulently mix the lobe and ambient medium?
\end{abstract}

\begin{keyword}
active galaxies \sep radio continuum \sep remnant galaxies
\end{keyword}

\end{frontmatter}



\section{Introduction}

The supermassive black holes (SMBHs) hosted by most galaxies \citep{Magorrian+1998, Haring+2004, Gultekin+2009} release vast amounts of energy through accretion, shaping both their host galaxies and the surrounding intergalactic medium. In the low-redshift universe, accretion most commonly powers relativistic jets that inflate synchrotron-emitting lobes, heating and displacing ambient gas \citep[e.g.,][]{Boehringer+1993, Carilli+1994, McNamara+2000, Fabian+2003}. This `jet-mode' feedback regulates star formation in massive galaxies \citep{Croton+2006} and suppresses cooling flows into cluster cores \citep[see e.g. reviews by][]{McNamara+2007, Fabian+2012}. Prescriptions for feedback depend on both the energetics and duration of the active phase, as well as the timescales over which the lobes interact with the surrounding gas \citep{Binney+2007, Alexander+2012, Turner+2015, Sullivan+2025}; e.g., through shock-heating driven by rapid expansion of the lobes \citep{Worrall+2012}, or the clearing of gas through the buoyant rise of lobes (or `bubbles') out of the cluster potential \citep{Churazov+2001}. Meanwhile, the length of the quiescent phase constrains how quickly the environment can return to equilibrium before the resumption of jet activity.

Radio-loud AGN activity is episodic \citep[as shown directly by {double-double radio galaxies};][]{Schoenmakers+2000}, with active phases lasting several tens of Myr before the jets switch off and the lobes begin to fade \citep{Parma+1999}. The remnant phase provides a means of constraining duty cycles through spectral \citep{Kardashev+1962, Murgia+2011, Turner+2018c, Quici+2022} or dynamical modelling \citep{Quici+2025}, and offers insight into the formation of cluster-scale radio structures such as halos, relics, and phoenixes \citep[e.g.,][]{Slee+2001, Ensslin+2002, vanWeeren+2009, deGasperin+2014, Shabala+2024}. Low-frequency surveys have, however, consistently found that remnants comprise only a small fraction of the radio population \citep[$<3$-5\%; e.g.,][]{Brienza+2017, Jurlin+2021, Quici+2021, Singh+2021, Dutta+2023}. 

\citet{Murgia+2011} investigated the large-scale environments  of `dying' radio galaxies in the B2 bright sample \citep[complete to 0.2-0.25~Jy at 408~MHz; see][]{Colla+1975} concluding that $\sim 86$\% of remnants are in clusters. Consequently, \citet{Murgia+2011} suggested radio emission
from the fading remnant lobes lasts longer in clusters than low-density environments as their expansion rate is greatly reduced (e.g., higher surface brightness and lower adiabatic losses). However, this argument is in conflict with dynamical modelling which predicts significantly higher radiative losses in remnants with a high jet power in their preceding active phase or (equivalently) expanding into a dense ambient medium \citep[][their Figure 4]{Turner+2018c}. Instead, it seems plausible the relatively high flux densities of radio sources in the B2 bright sample bias the selection against all but the brightest remnant lobes; i.e., high jet power, dense ambient medium, or very-low redshift \citep[e.g.,][]{Turner+2023a}.


Recent studies utilising deep 150~MHz observations from the \textit{Low Frequency Array} (LOFAR), supplemented by 325~MHz and 1.4~GHz observations from the \textit{Giant Meterwave Radio Telescope} (GMRT) and \textit{Jansky Very Large Array} (VLA), are now finding a deficit of remnants in massive clusters \citep{Jurlin+2021, Singh+2021, Dutta+2023}. Specifically, \citet{Jurlin+2021} found only 3/13 (23\%) of confirmed remnants in the well-studied Lockman Hole reside in cluster environments. \citet{Singh+2021} and \citet{Dutta+2023} meanwhile complied samples of remnants in the XMM-LLS field with flux densities reaching down to 1~mJy at 150~MHz and 6~mJy at 325~MHz, respectively. These authors found only 2/15 (13\%) and 1/15 (7\%) of remnants inhabit massive clusters. These findings suggest that the faint remnant lobes of `dying' AGNs in poorer group environments are either more ubiquitous or long-lived than those in clusters.

We propose that the observed scarcity of remnant AGNs in massive clusters instead results (at least in part) from the dynamics of the jet-inflated lobes. Specifically, a rapid `implosion' occurs upon the cessation of jet activity for lobes residing in a dense ambient medium, or with a weak jet in the preceding active phase. This `lobed' morphology is associated with both classes of the canonical \citet{Fanaroff+1974} dichotomy, however, the low-powered FR-I class includes a sizeable fraction of `tailed' sources; e.g., 459/1256 (37\%) of FR-Is \citep{Mingo+2019}. The (active) tailed population has a modest preference to reside in cluster environments (48-49\% compared to 20-35\% for radio AGNs with similar luminosities to confirmed remnants; \citealt{Mingo+2019}; \citealt{Croston+2019}, their Figure 3). 
However, such objects are atypical of most confirmed remnants \citep[e.g.,][]{Jurlin+2021, Quici+2021} suggesting either that remnant tailed sources are difficult to detect or there is a bias against their selection as remnant candidates (e.g., their core-dominated spectra may not meet spectral curvature or ultra-steep spectrum criterion; \citealt{Turner+2018a}, their Figure 13; \citealt{Stewart+2026}). These factors may contribute to a bias against observing remnants in massive clusters but cannot on their own explain the large deficit seen in recent observations. As such, we do not explicitly consider tailed radio sources further in this work.

The paper is structured as follows. In \cref{sec:lobe dynamical model}, we present the key equations describing the lobe dynamics in both the active and remnant phase, and perform asymptotic analysis to understand potential lobe behaviour in regions of the jet power--ambient medium parameter space. In \cref{sec:results}, we derive the `implosion' timescale as a function of jet power and the ambient medium and compare to the timescale over which the remnant is observable; we discuss the expected bias against detecting lobed remnants in massive clusters. Then, in \cref{sec:discussion}, we discuss the limitations of the analytical model, introduce buoyancy and a stabilising magnetic sheath, and compare our predictions to a hydrodynamic simulation. We conclude in \cref{sec:conclusion} relating the implications of our findings to the observed energy budget of AGN feedback.

\section{Lobe Dynamical Model}
\label{sec:lobe dynamical model}

\begin{figure*}
\centering
\includegraphics[width=0.625\textwidth,trim={0 0 0 0},clip]{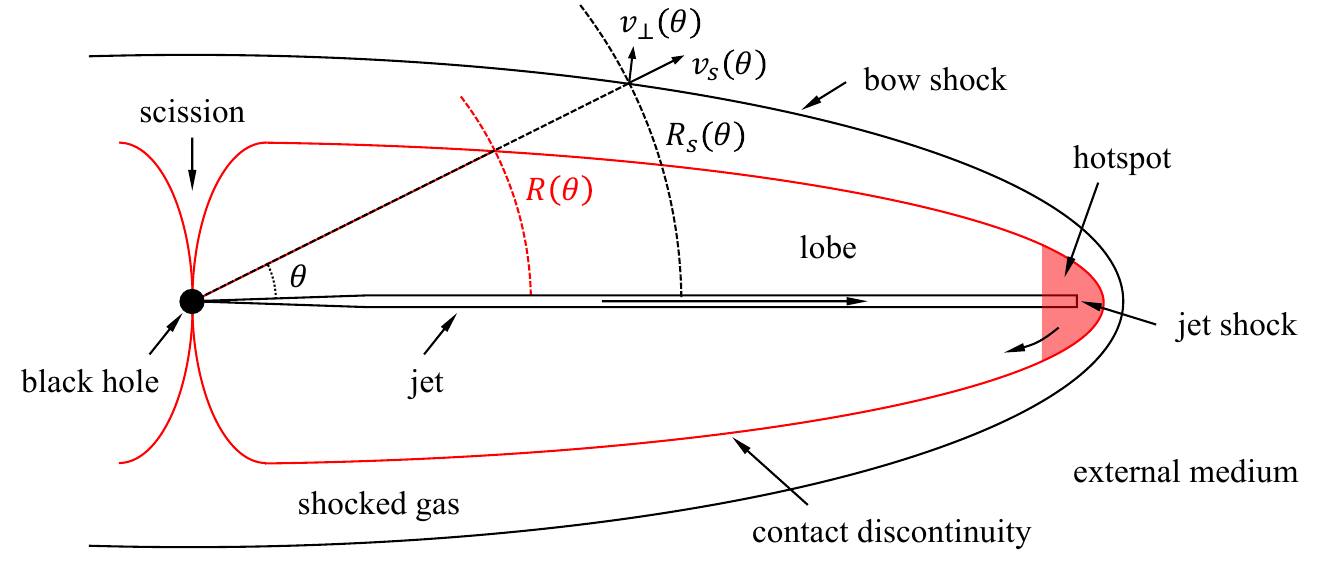}
\caption{Schematic of the \citet{Turner+2023a} dynamical model for the lobe and shocked shell (\cref{sec:pressure-driven expansion}), showing the scission of the two lobes immediately upon the cessation of jet activity (\cref{sec:buoyancy-driven expansion}).}
\label{fig:schematic}
\end{figure*}

We explore the behaviour of low-powered remnant lobes using the \textit{Radio AGN in Semi-analytic Environments} (RAiSE) model, as described by \citet{Turner+2023a}.
Their dynamical model assumes a high-powered relativistic plasma jet drills through the ambient medium generating a bow shock which radiates outwards from the jet-head. The plasma in the jet is forced backwards towards the active nucleus by the pressure of the shocked ambient gas inside the bow shock; the region filled by the shock-accelerated plasma is referred to as the lobe (see \cref{fig:schematic}).

We briefly introduce key equations describing the pressure-driven expansion of the lobe and shocked gas shell in the active and remnant phases (\cref{sec:pressure-driven expansion}), then investigate the limiting behaviour of these equations for remnant lobes using asymptotic analysis (\cref{sec:asymptotics}). We discuss which regions of parameter space (e.g., jet power, ambient gas density) lead to the predicted behaviour of each asymptotic limit: (1) a reduced growth rate relative to the active phase, or (2) a rapid implosion of the remnant lobe to zero radius (\cref{sec:radio source parameters that lead to implosion}).


\subsection{Pressure-driven expansion}
\label{sec:pressure-driven expansion}

The expansion of the relativistic jet and shocked gas shell is modelled by \citet{Turner+2023a} considering a two-phase fluid with momentum flux contributions from both the jet and thermal shocked gas (see their Section 2). 
We focus on their equations describing the later-time evolution of the shocked shell in this summary. {In their framework, the origin of the coordinate system is centred on the black hole with the jet aligned along the $\theta = 0$ radial line (see \cref{fig:schematic}), with assumed azimuthal symmetry in the lobe and shocked shell, and spherical symmetry in the ambient medium.}
The RAiSE model solves a system of ordinary differential equations (ODEs; derivatives with respect to the source age, $t$) for the radius, velocity and Lorentz factor at the surface of the shocked shell as a function of the polar angle $\theta$ from the jet axis, i.e., $R_s(t, \theta)$, $v_s(t, \theta)$ and $\gamma_s(t, \theta)$, respectively. \citet{Turner+2023a} adopt a numerical scheme using a fourth-order Runge-Kutta method in terms of a system of three non-linear first order ODEs for each polar angle.


The following system of equations must be solved for each small angular element $[\theta - \delta\theta/2, \theta + \delta\theta/2)$ of the shocked shell \citep[see][their Equation 19]{Turner+2023a}:
\begin{equation}
\begin{split}
\dot{R}_s(t, \theta) &= v_s \\
\dot{v}_s(t, \theta) &= \frac{3\Gamma_{\rm c} - \beta}{2 R_s [1 + (\gamma_s v_s/c)^2] [\zeta_{\rm s} \gamma_s/\eta_{\rm s}]^2} \\
&\quad\quad \times 
\left[\!\;\frac{3 (\Gamma_{\rm c} - 1) Q R_s^{\!\;\beta - 2} \eta_{\rm s}^{3 - \beta}\zeta_{\rm s}^{2}}{2 \pi (3\Gamma_{\rm c} - \beta)I\!\!\!\;\; v_s (\rho a^\beta)} - [\zeta_{\rm s} \gamma_s v_s/\eta_{\rm s}]^2 - c_{\rm x}^2/\Gamma_{\rm c} \;\!\right]
\\
\dot{\gamma}_s(t, \theta) &= \frac{\gamma_s^3 v_s \dot{v}_s}{c^2} ,
\end{split}
\label{supersonic system}
\end{equation}
where $Q$ is the instantaneous jet power (each jet; i.e., $Q_\text{tot} = 2Q$), $c$ is the speed of light, $c_{\rm x}$ is the sound speed of the ambient medium, and $\Gamma_{\rm c} \approx \tfrac{5}{3}$ is the adiabatic index of the lobe plasma and shocked gas. The ambient gas density is locally described by a power law of the form $\rho_{\rm x}(r) = \rho [r/a]^{-\beta}$; the RAiSE framework models complex gas density profiles as a continuous function that combines numerous such power laws each defined over small radial intervals. The variable $I$ is a source-specific constant that represents the steady state distribution of energy throughout the lobe and shocked shell by the jet. That is,
\begin{equation}
I = \int_0^{\tfrac{\pi}{2}} {\eta_{\rm s}}^{3 - \beta}(\theta')\!\; {\zeta_{\rm s}}^2(\theta') \sin\theta' \delta\theta' ,
\label{I integral}
\end{equation}
where $\eta_s$ and $\zeta_s$ are geometric factors defined in Equations 12 and 13 of \citet{Turner+2023a}, respectively. Both these factors converge to unity in the limiting case of a spherical radio source with $A = A_s = 1$, where $A$ and $A_s$ are the late-time axis ratios of the lobe and shocked gas shell, respectively; i.e., $\eta_s = \zeta_s = I = 1$ for a spherical lobe and shocked shell.

\subsection{Asymptotic analysis of remnant dynamics}
\label{sec:asymptotics}

We can greatly simplify the expression for the acceleration in our system of differential equations under the assumption of a spherical radio source (i.e., $A_s \rightarrow 1$) expanding at non-relativistic velocities (i.e., $\gamma_s \rightarrow 1$ and $\Gamma_{\rm c} \rightarrow \tfrac{5}{3}$) as follows:
\begin{equation}
\begin{split}
\dot{v}_s &= \frac{5 - \beta}{2 R_s}\left[\!\;\frac{Q R_s^{\!\;\beta - 2}}{(5 - \beta)\pi v_s (\rho a^\beta)} - v_s^2 - \tfrac{3}{5}c_{\rm x}^2 \;\!\right] ,
\end{split}
\label{simplified system}
\end{equation}
where $\dot{v}_s < 0$ for all $t > 0$, and for active sources (i.e., $Q = Q_\text{on} > 0$, for constant $Q_\text{on}$) it can be shown that $v_s > 0$.


We investigate the behaviour of the radio lobes in a remnant phase (i.e., $Q = 0$) for two limiting cases: (1) $v_s \gg c_{\rm x}$, momentum driven expansion (i.e., momentum of the lobe and shocked shell after the jet ceases); and (2) $v_s \ll c_{\rm x}$, ambient pressure confined expansion.

\subsubsection{Momentum driven expansion}

The expression for the acceleration of the remnant-phase lobe in this limiting case is as follows:
\begin{equation}
\begin{split}
\dot{v}_s &= -\frac{(5 - \beta)v_s^2}{2 R_s} .
\end{split}
\label{asymptote 1}
\end{equation}
The solution to this simplified equation, in terms of the shocked shell radius $R_s$, is given by,
\begin{equation}
\begin{split}
R_s(t) &= R_\star \left[\frac{(7 - \beta)(t - t_\star)}{2} \right]^{2/(7 - \beta)} ,
\end{split}
\label{solution 1}
\end{equation}
where $R_\star$ and $t_\star$ are the radius and time offset that ensure continuity of the lobe radius and its velocity upon the cessation of jet activity (i.e., at $t = t_\text{on}$, where $t_\text{on}$ is the duration of the active phase). This is a slowly increasing, non-linear function of time for $t \gg t_\star$; for a constant density ambient medium, $R_s(t) \propto t^{2/7}$, ensuring moderate growth throughout the remnant phase.

\subsubsection{Ambient pressure confined expansion}

The expression for the acceleration of the remnant-phase lobe in this second limiting case is given by,
\begin{equation}
\begin{split}
\dot{v}_s &= -\frac{3(5 - \beta)c_{\rm x}^2}{10 R_s} ,
\end{split}
\label{asymptote 2}
\end{equation}
which has a rather inelegant solution, in terms of the shocked shell radius $R_s$, as follows:
\begin{equation}
\begin{split}
R_s(t) &= R_\star \;\!e^{- \,\left(\text{\,erf\,}^{-1}\!\!\sqrt{\frac{3(5 - \beta)c_{\rm x}^2}{5\pi R_\star^2}(t - t_\star)^2 }\,\right)^2} \quad \text{ if }|t - t_\star| \leqslant t_{\rm crit} ,
\end{split}
\label{solution 2}
\end{equation}
where $\text{\,erf\,}^{-1\!}$ is the inverse error function, and $R_\star$ and $t_\star$ have the same meaning as in \cref{solution 1}, albeit with different values. This solution is valid on a restricted domain defined by a critical timescale, $t_{\rm crit}$, given by,
\begin{equation}
\begin{split}
t_{\rm crit} &= \frac{R_\star}{c_{\rm x}} \sqrt{\frac{5\pi}{3(5 - \beta)}} .
\end{split}
\label{implosion timescale}
\end{equation}
The solution approaches $R_s \rightarrow 0$ when $|t - t_\star| = t_{\rm crit}$, noting the inverse error function has the limiting value $\text{\,erf\,}^{-1}(1) \rightarrow \infty$. That is, the lobe `implodes' to zero radius in a finite time $t_{\rm crit}$ upon the cessation of the jet; for a constant density ambient medium, $t_{\rm crit} \approx R_\star/c_{\rm x}$, implying shorter implosion timescales for more compact radio lobes ({and} which also satisfy the asymptotic condition $v_s \ll c_{\rm x}$). For example, a relatively compact radio lobe with $R_\star \approx 10$\,kpc and $c_{\rm x} \approx 0.01c$ has an implosion timescale of $t_{\rm crit} \approx 3.3$\,Myr.

We perform a Taylor series expansion at $t = t_\star$ to interpret the behaviour of this solution within the restricted domain, giving:
\begin{equation}
\begin{split}
R_s(t) &= R_\star - \frac{3(5 - \beta)c_{\rm x}^2}{20 R_\star}(t - t_\star)^2 + \mathcal{O}((t - t_\star)^4) .
\end{split}
\label{solution 2 taylor}
\end{equation}
This solution takes the form of an inverted parabola, implying the lobe decreases in radius slowly at first before rapidly `imploding' to a radius of $R_s = 0$.

\subsection{Radio source parameters that lead to implosion}
\label{sec:radio source parameters that lead to implosion}

We study the sets of radio source parameters that yield the first limiting case, $v_s \gg c_{\rm x}$, in order to place a bound on the possible parameters that lead to imploding remnant lobes. This necessitates deriving a solution for the velocity of the shocked shell in the active phase, $v_s$, in order to relate this variable to (e.g.) the jet power and properties of the ambient medium.

The solution to \cref{simplified system} in the limit $v_s \gg c_{\rm x}$, in terms of the shocked shell radius{ at the cessation of jet activity}, $R_s(t_\text{on})$, is given by,
\begin{equation}
\begin{split}
R_s(t_\text{on}) &= \left[ \frac{(5 - \beta)^3 Q_\text{on}}{9\pi(11 - \beta)(\rho a^\beta)} \right]^{1/(5 - \beta)} t_\text{on}^{3/(5 - \beta)} ,
\end{split}
\label{solution 3}
\end{equation}
which yields the following expression for the velocity of the shocked shell {at the end of} the active phase:
\begin{equation}
\begin{split}
v_s(t_\text{on}) &= \frac{3}{5 - \beta} \left[ \frac{(5 - \beta)^3 Q_\text{on}}{9\pi(11 - \beta)(\rho a^\beta)} \right]^{1/(5 - \beta)} t_\text{on}^{(\beta - 2)/(5 - \beta)} .
\end{split}
\label{solution 4}
\end{equation}

The first asymptotic solution is therefore valid under the following condition, derived using \cref{solution 4} and the standard expression for the sound speed, $c_{\rm x} = \sqrt{\Gamma_{\rm c}k_{\rm B} T/\bar{m}}${, where $\bar{m}$ is the average mass of plasma particles in the ambient medium.} That is,
%
%
\begin{equation}
\begin{split}
&\left[\frac{Q_\text{on}}{10^{35.5}\,\rm W} \right]\!\;\! \left[\frac{\rho(a = 10\rm\,kpc)}{5\times 10^{-24}\,\rm kg\, m^{-3}} \right]^{-1\!\;\!} \left[\frac{T}{10^7\rm\, K} \right]^{(\beta - 5)/2} \left[\frac{t_\text{on}}{10\rm\, Myr} \right]^{\!\;\beta - 2} \\
&\qquad\quad \gg 0.0806 (11 - \beta) \big[\!\; 0.1264 (5 - \beta) \big]^{2 - \beta} \ \ \ \approx 1,
\end{split}
\label{critical line}
\end{equation}
where the right-hand side of the inequality is bounded in the range $[0.35, 0.73]$ for plausible ambient density profiles \citep[i.e., $0 \leqslant \beta < 2$; cf.][]{Turner+2015}. {Our choice of ambient gas density (at $a= 10$\;kpc) and temperature correspond to a typical cluster with halo mass $M_\text{halo} \approx 10^{13.63}$\;$\text{M}_\odot$ (see \ref{sec:relation to large-scale host cluster parameters}). The shocked shell has a radius in the range $[4.22, 7.78]$~kpc for this ambient medium with a $Q_\text{on} = 10^{35.5}$\;W jet power and $t_\text{on} = 10$\;Myr active age.}

This analytic approach predicts the upper bound {(for some interpretation of `much larger than')} in parameters that lead to imploding remnant lobes is a linear function (in log-space) of the jet power and both the density and temperature of the ambient medium, with a positive correlation between these variables.



\section{Results}
\label{sec:results}

The predicted dichotomy in remnant evolution as a function of location in the jet power--ambient medium parameter space is investigated using the complete form of the differential equations (cf. \cref{supersonic system}), as captured in the RAiSE dynamical model \citep{Turner+2023a}. We encapsulate the ambient density and temperature, in addition to the slope of the density profile $\beta$, through a single parameter: the group/cluster halo mass, $M_{\rm halo}$. The shape of the density profile is informed by X-ray observations while the virial radius (required to scale the profile) and temperature are related to the halo mass based on semi-analytic galaxy evolution models \citep[for details, see][]{Turner+2015}.

We model the remnant implosion timescale as a function of the active jet power, lifetime and properties of the ambient medium in \cref{sec:remnant implosion timescales}, and compare the results of the full dynamical model to the log-linear upper bound for implosion predicted by the asymptotic analysis. We compare the implosion timescale to the synchrotron cooling timescale and discuss the potential importance of remnant `implosion' on population statistics (\cref{sec:remnant visibility}).

\subsection{Remnant implosion timescales}
\label{sec:remnant implosion timescales}

We calculate the implosion timescale, $t_{\rm crit}$, for remnant lobes with a range of active ages, $t_{\rm on}$, across the jet power--ambient medium $(Q_\text{tot}, M_{\rm halo})$ parameter space; default model parameters (e.g., ratio of the radius of jet spine to jet sheath) are assumed in the RAiSE dynamical model based on their calibration to hydrodynamic simulations \citep[see][]{Turner+2023a}. Notably, the late-time lobe axis ratio takes a value of $A=2.83$, in contrast to the simplifying assumptions of the asymptotic analysis (cf. \cref{sec:asymptotics}). The inverse-Compton radiative losses and properties of the semi-analytic environments are calculated for redshift $z=0.1$.

\begin{figure*}
\centering
\includegraphics[width=\textwidth,trim={95 0 195 35},clip]{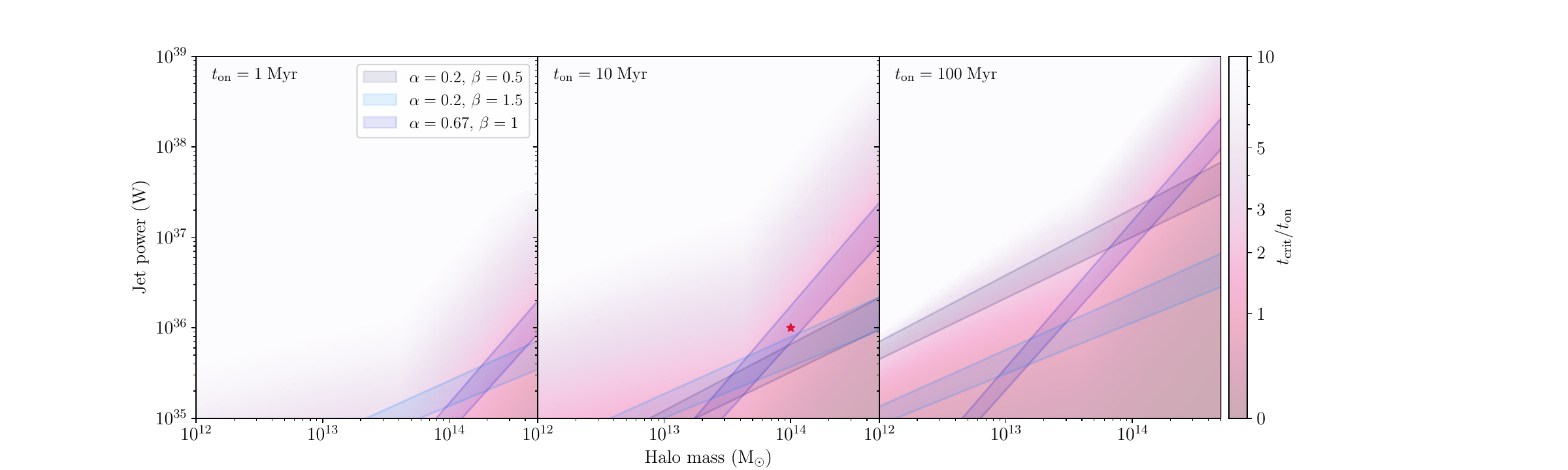}
\caption{The implosion timescale ($t_{\rm crit}/t_\text{on}$) derived using the RAiSE dynamical model for remnant lobes across the jet power--ambient medium $(Q_\text{tot}, M_{\rm halo})$ parameter space.
The implosion timescale is shown in red/pink shading for three active ages: 1, 10 and 100~Myr. The asymptotic analysis upper bound for parameters that lead to implosion is shown ({blue} shaded lines) for comparison assuming plausible values of the slopes of the approximating power laws, $\alpha$ and $\beta$. The red star marks the location in parameter space assessed against a hydrodynamic simulation in \cref{sec:diffusion and fluid instabilities}.}
\label{fig:tcrit}
\end{figure*}

The implosion timescale is plotted in \cref{fig:tcrit} for three active ages: 1, 10 and 100~Myr. The shortest implosion timescales (relative to the active age; i.e., $t_\text{crit}/t_\text{on}$) occur for low-powered sources in the densest cluster environments{; in fact, their rate of implosion (at velocity $\approx c_{\rm x}$; cf. \cref{implosion timescale}) is faster than the average expansion rate across their active phase (i.e., subsonic for most of their evolution, $v_s < c_{\rm x}$).} The implosion timescale does not increase in direct proportion to the active age leading to fractional smaller implosion timescales for the 100~Myr active age. Consequently, few large remnant radio lobes would be expected in observed populations except in relatively poor environments (e.g., $M_\text{halo}<10^{13.5}$~M$_\odot$), or with very-high jet powers their preceding active phase (e.g., $Q_\text{tot} \gg 10^{38}$~W).

We compare predictions of the upper bounds from the asymptotic analysis (\cref{sec:radio source parameters that lead to implosion}) to the full dynamical model. However, we must convert the log-linear relationship presented in \cref{sec:radio source parameters that lead to implosion} from functions of the ambient gas density and temperature to a function of the cluster mass to directly compare with RAiSE; the cluster mass is additionally a more readily constrained parameter in observed populations. That is, the upper bound {(again, for some interpretation of `much larger than')} for the first asymptotic solution (\cref{critical line}) becomes:
\begin{equation}
\begin{split}
&\left[\frac{Q_\text{on}}{10^{35.5}\,\rm W} \right]\!\;\! \left[\frac{M_\text{halo}}{10^{13.63}\rm\, M_\odot} \right]^{f(\alpha,\,\beta,\,\beta')} \left[\frac{t_\text{on}}{10\rm\, Myr} \right]^{\!\;\beta - 2} \\
&\qquad\quad \gg 0.0139^{1.526 - \beta'\!} (3 - \beta') (11 - \beta) \big[\!\; 0.1264 (5 - \beta) \big]^{2 - \beta},
\end{split}
\label{critical line halo}
\end{equation}
where $f(\alpha,\beta,\beta') = [3\alpha(\beta - 5) - 2\beta']/6$, and $\alpha \in [0, 0.7)$, $\beta \in [0, 2)$ and $\beta' = 0.76 \pm 0.09$ are the exponents of approximating power laws for properties of the ambient medium; see \ref{sec:relation to large-scale host cluster parameters} for the derivation of \cref{critical line halo} and a complete description of the simplifying assumptions. Briefly, the halo mass--temperature relationship is parameterised through $\alpha$ \citep[i.e., $\alpha = 2/3$ for massive clusters, flattening for poorer groups; e.g.,][]{Sun+2009}, and the local slope of the ambient gas density profile is modelled with $\beta$ as before (i.e., $\beta \gtrsim 0$ for small, young lobes, and $\beta \lesssim 2$ for large, old lobes).

The log-linear asymptotic relationship is shown in \cref{fig:tcrit} assuming three sets of plausible values for the power law exponents $\alpha$ and $\beta$: $(\alpha, \beta) = (0.2, 0.5)$, $(0.2, 1.5)$ and $(0.67, 1)$. {We present the upper bound assuming equality in the asymptotic relationship (i.e., a very tight upper bound).} The width of the lines corresponds to the measurement uncertainty in $\beta'$ based on observed clusters (see \ref{sec:relation to large-scale host cluster parameters}). The asymptotic solutions are (unsurprisingly) consistent with the predictions of the full dynamical model. This result gives confidence that the RAiSE model behaves as expected given the comparative simplicity of the asymptotic analysis.

\subsection{Remnant visibility}
\label{sec:remnant visibility}

The synchrotron emission from remnant bubbles will cease upon implosion, potentially explaining the scarcity of low-powered remnants in radio surveys. However, we need to compare the implosion timescale to that of synchrotron cooling in stable lobes to assess the significance of `implosions' on population statistics; i.e., are most remnants already undetectable at radio frequencies prior to an implosion event?

\begin{figure*}
\centering
\includegraphics[width=\textwidth,trim={95 0 195 35},clip]{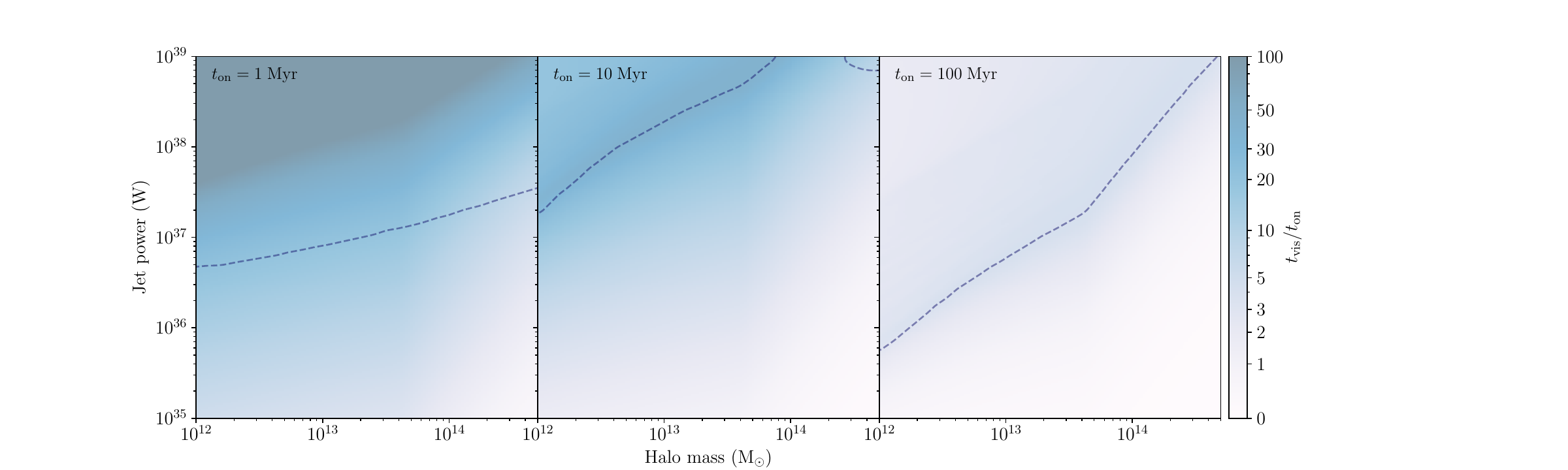}
\caption{The visible timescale ($t_{\rm vis}/t_\text{on}$) at 150~MHz derived using the RAiSE dynamical model for remnant lobes across the jet power--ambient medium $(Q_\text{tot}, M_{\rm halo})$ parameter space.
The visible timescale is shown in blue shading for three active ages: 1, 10 and 100~Myr. {The regions of parameter space below the blue line have a synchrotron cooling timescale comparable to (or longer than) the implosion timescale.}}
\label{fig:visible}
\end{figure*}

\begin{figure}
\centering
\includegraphics[width=\columnwidth,trim={15 7 35 40},clip]{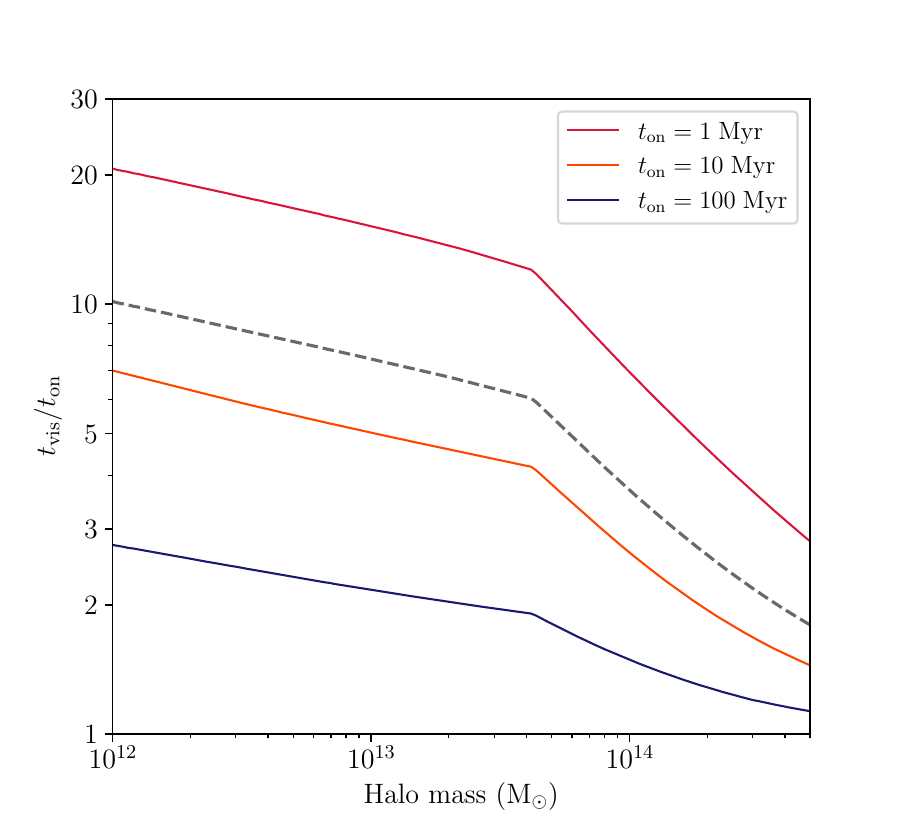}
\caption{The mean visible timescale (at 150~MHz) of remnant lobes expected across an observed population as a function of cluster mass. The occurrence of each jet power is weighted as $p(Q) \propto Q^{-1.5}$ following \citet{Quici+2025}. The visible timescale is shown for three active ages: 1, 10 and $100$~Myr in red, orange and blue, respectively. The occurrence of each active age is additionally weighted as $p(t_\text{on}) \propto t_\text{on}^{-1}$ \citep[cf.][]{Quici+2025} to give a prediction for the overall population (grey dashed line).}
\label{fig:hist}
\end{figure}

We derive the timescale, $t_\text{vis}$, over which some non-zero radio-frequency emission remains at 150~MHz observer-frame for redshift $z=0.1$; see \citet[][their Section~3]{Turner+2023a} for a complete description of the synchrotron emission calculation in RAiSE. The visible timescale cannot be longer than the implosion timescale as no emission remains upon implosion. We plot this visible timescale in \cref{fig:visible} for three active ages{; regions of parameter space below the blue line have their emission dynamically curtailed by the implosion}. As expected, remnant lobes with a $1$~Myr active age are visible for a few tens of Myr when stable to implosion; however, those that are unstable to implosio{(based on the full RAiSE model) have their emission dynamically curtailed within} under 1~Myr. This leads to a scarcity of lobed remnants in dense clusters compared to poorer groups (e.g., $M_\text{halo} < 10^{13.5}$~M$_\odot$ versus $M_\text{halo} \gtrsim 10^{13.5}$~M$_\odot$). This is explored further in \cref{fig:hist} which shows the expected remnant counts (i.e., counts are proportional to the visible timescale for a given active age) as a function of cluster mass for a power law jet power function of the form $p(Q) \propto Q^{-1.5}$ \citep[cf.][]{Quici+2025}. The number of observable remnants in massive clusters is strongly suppressed by this hypothesised implosion mechanism.

The visible timescale exhibits similar behaviour for remnants with a preceding 10~Myr active phase. By contrast, synchrotron cooling for the remnants with a 100~Myr active age is {often faster than} the implosion timescale across much of the jet power--ambient medium parameter space {(notably towards the top-left of 100~Myr panel in \cref{fig:visible})}. That is, the large sizes reached after a 100~Myr active age greatly increase the implosion timescale, whilst the larger number of injected particles increases their susceptibility to inverse-Compton radiative losses. {The tension between rapid synchrotron cooling (top-left) and short implosion timescales (bottom-right) leads to a} somewhat linear region through the centre of the parameter space with visible timescales reaching up to few tens of Myr. {Regardless, the rapid implosion of the more prevalent low-powered sources} again leads to a relative scarcity of lobed remnants in massive clusters (\cref{fig:hist}).

The relative ubiquity of short-lived radio sources gives greater importance to our findings for the 1~Myr active age; i.e., $p(t_\text{on}) \propto t_\text{on}^{-1}$ \citep[cf.][]{Quici+2025}. We combine our results for each active age by weighting their visible timescale with this function, as shown in \cref{fig:hist} by the grey dashed line. These results show that remnant lobes in the most massive clusters ($M_\text{halo} \sim 10^{14.5}$~M$_\odot$) will be under-counted by a factor of five compared to poor groups ($M_\text{halo} \sim 10^{12}$~M$_\odot$). The absolute counts in observed populations will of course also depend on the number of active jets in these cluster environments. \citet{Croston+2019} find 30\% of active radio AGNs with $L_{150\text{ MHz}} > 10^{26}$~W\,Hz$^{-1}$ (typical of confirmed remnants) reside in catalogued clusters (complete for $M_\text{halo} > 10^{14}$~M$_\odot$). The factor of a few under-counting we predict in this mass range is sufficient to reduce observed remnant counts in massive clusters down to the 7-13\% range of recent observations \citep[cf.][]{Singh+2021, Dutta+2023}.

\section{Discussion}
\label{sec:discussion}

The accuracy of RAiSE dynamical model predictions has been assessed in comparison to hydrodynamic simulations throughout the jet/lobe evolutionary history \citep[][their Figures 3-5]{Turner+2023a}, and other analytical models for the late-time lobe expansion \citep{Turner+2023b}. We discuss the potential limitations of this dynamical model in predicting the implosion of remnant lobes; in particular, buoyancy acting on the underdense lobe plasma (\cref{sec:buoyancy-driven expansion}), magnetic suppression of fluid instabilities at the interface of the lobe and shocked shell (\cref{sec:magnetic field surface tension}), and turbulent mixing of lobe plasma and shocked gas by Rayleigh-Taylor and Kelvin-Helmholtz fluid instabilities (\cref{sec:diffusion and fluid instabilities}).

\subsection{Buoyancy-driven acceleration}
\label{sec:buoyancy-driven expansion}

The underdense plasma within the lobe {(and the overdense plasma in the shocked gas shell)} will lead to an additional buoyant acceleration term along the jet axis. This is difficult to model in detail within the RAiSE framework, but an upper bound can be captured by assuming the scission of the two lobes immediately upon the cessation of the jet (see \cref{fig:schematic}). We consider the buoyant force exerted along the jet axis for each half of the lobe based on the density and gravitational acceleration of the gas in its hydrostatic equilibrium state.


The local gravitational acceleration, $g(r)$, acting on the ambient gas is related to the equilibrium gas density profile, $\rho_{\rm x}(r)$, as follows:
\begin{equation}
g(r) = -\frac{1}{\rho_{\rm x}(r)}\frac{\partial p_{\rm x}(r)}{\partial r} \ \ \ = \frac{\beta(r) c^2_{\rm x}}{r\Gamma_{\rm c}},
\label{gravity}
\end{equation}
where $p_{\rm x}(r)$ is the ambient pressure profile, related to the density under the assumption of an approximately isothermal gas.

The mass of plasma within the bubble (or equivalently one-half of the lobe after scission through the core) is related to the injected energy as follows:
\begin{equation}
m_\text{bubble} = \frac{Q_\text{on} t_\text{on}}{(\bar{\gamma}_j - 1)c^2},
\end{equation}
where $Q_\text{on}$ and $t_\text{on}$ are the jet power and duration of the active phase, and $\bar{\gamma}_j$ is the Lorentz factor of the bulk flow in the jet \citep[][their Equation 9]{Turner+2023a}. {Meanwhile, the mass of ambient plasma swept up into the dense shocked gas shell is given by,}
\begin{equation}
\begin{split}
m_\text{shock} &= 2\pi \int_0^{\tfrac{\pi}{2}} \!\!\int_0^{R_s(\theta')} 
\!\rho_{\rm x}(r)\,r^2dr\sin\theta'\delta\theta' ,
\end{split}
\end{equation}
{compared to the lower mass in that region for gas in the hydrostatic equilibrium state. That is,}
\begin{equation}
\begin{split}
m_\text{equil} &= 2\pi \int_0^{\tfrac{\pi}{2}} \!\!\int_{R(\theta')}^{R_s(\theta')} 
\!\rho_{\rm x}(r)\,r^2dr\sin\theta'\delta\theta' ,
\end{split}
\end{equation}
where $R(\theta)$ is the radius of the lobe at polar angle $\theta$. The shape of the lobe is explicitly modelled by \citet[][their Sections 2.3 to 2.4]{Turner+2023a}; i.e., the late-time limit approaches $R(\theta) = b\eta_s(\theta) R_s(\theta)/\eta(\theta)$, where $b\approx 1.07$ is the ratio of the
shocked shell to lobe radii along the jet axis, and $\eta(\theta)$ is a geometric factor defined in Equation 14 of \citet{Turner+2023a}.

{The lobe and shocked gas shell continue to act as a coupled system immediately after the jet switches off, with a net acceleration along the jet axis (i.e., other components are balanced by symmetry). At later times, the lobe bubble may begin to propagate through the shell with the two components increasingly behaving as separate entities, leading to more complicated dynamics. We therefore consider the buoyant acceleration for the former case (applicable early in the remnant phase) with contributions from both the lobe and shocked gas shell. That is},
\begin{equation}
\begin{split}
a_\text{buoy}(\theta &= 0) \approx \frac{2\pi c_{\rm x}^2}{\Gamma_{\rm c} (m_\text{bubble} + m_\text{shock})}
\int_0^{\tfrac{\pi}{2}} \Bigg[\!\;\!\int_0^{R(\theta')} 
\!\rho_{\rm x}(r)\beta(r)\,r dr \\&+ \int_{R(\theta')}^{R_s(\theta')} 
\!\rho_{\rm x}(r)\, \bigg(1 - \frac{m_\text{shock}}{m_\text{equil}}\bigg)\,\beta(r)\,r dr \;\! \Bigg]\;\! \cos\theta'\sin\theta'\delta\theta' ,
\end{split}
\label{buoyancy}
\end{equation}
{where the second integral assumes the dense gas in the shocked shell follows the same profile as the undisturbed ambient medium, but with increased densities.}
%
%
The component of this acceleration vector acting on each small angular element of the surface of the shocked shell {(as required by RAiSE) is obtained by projecting the acceleration in \cref{buoyancy} as follows:}
\begin{equation}
\begin{split}
\dot{v}_{s,\text{buoy}}(\theta) = a_\text{buoy}(\theta = 0) \cos\theta ,
\end{split}
\label{buoyant_comp}
\end{equation}
where this expression is only accurate for angles $\theta$ that remain on the top-surface of the bubble (i.e., we do not consider the bottom surface explicitly). Consequently, the addition of buoyancy to the model in this manner will initially be accurate, but slowly diverges from physical expectation as the bubble deforms due to fluid instabilities {and begins to propagate through the shocked gas shell (if it has not already imploded)}.

\begin{figure*}
\centering
\includegraphics[width=\textwidth,trim={95 0 200 35},clip]{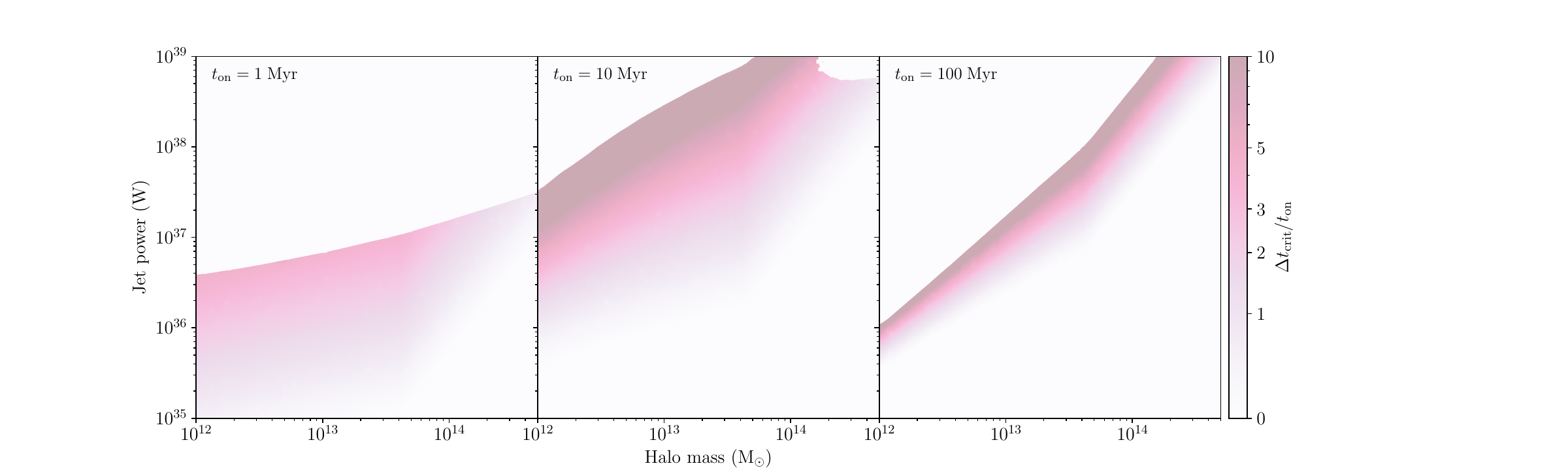}
\caption{Comparison of the implosion timescale derived using the RAiSE dynamical model either including or excluding a buoyant force along the jet axis. The difference in the implosion timescales ($\Delta t_{\rm crit}/t_\text{on}$; longer timescale with buoyancy) is shown for remnant lobes across the jet power--ambient medium $(Q_\text{tot}, M_{\rm halo})$ parameter space; see \cref{fig:tcrit} for complete description. The pixelation in the centre panel results from numerical errors in regions of parameter space that are quasi-stable to implosion.}
\label{fig:buoyancy}
\end{figure*}

The implosion timescales for dynamical models including or excluding a buoyant acceleration term are compared in \cref{fig:buoyancy}. These results show that the buoyant force is too weak to prevent the implosion of most remnant lobes; only those with long implosion timescales (i.e., marginally unstable to implosion) can be stabilised with the addition of an outward buoyant force. The results presented in this work include the buoyant force for completeness (i.e., \cref{fig:tcrit,fig:visible,fig:hist}).

\subsection{Magnetic suppression of fluid instabilities}
\label{sec:magnetic field surface tension}

The lobes of observed radio sources are known to be enclosed by a thin sheath with relatively strong tangential magnetic fields. This sheath provides an effective surface tension that acts to suppress fluid instabilities along the interface between the lobe and shocked gas shell. We investigate the suppression of Rayleigh-Taylor instabilities in detail, but note that similar arguments and conclusions apply for Kelvin-Helmholtz instabilities.

The magnetic field strength in the sheath enclosing the lobe is related to the pressure in the lobe and the equipartition factor, $q \sim 0.03$ \citep[cf.][]{Turner+2018b}, as follows:
\begin{equation}
B = \sqrt{\frac{2\mu_0 q p(r)}{(\Gamma_{\rm c} - 1)(q + 1)}}.
\label{B field}
\end{equation}
where $\mu_0$ is the vacuum permeability. The magnetic field strength is known to be highest in the sheath, so larger equipartition factors are expected compared to the lobe (e.g., $q \approx 1$ in equipartition).

The critical magnetic field strength required for the sheath to suppress the longest Rayleigh-Taylor mode, $\lambda$, is given by,
\begin{equation}
B_\text{crit} = \sqrt{\frac{\;\!\mu_0 g(r) \Delta \rho(r) \;\! \lambda}{2\pi}},
\end{equation}
where $g(r)$ is the local gravitational acceleration (\cref{gravity}), and $\Delta\rho$ is the density contrast between the shocked gas shell and the lobe. For low-powered radio sources we can reasonably assume the lobe and shocked shell are in pressure equilibrium with the ambient medium; i.e., $p \sim p_{\rm x}$ and $c_{\rm x}^2 \sim \Gamma_{\rm c} p /\rho_{\rm x}$. Similarly, we can assume the density of the shocked gas shell, $\rho_s$, is much greater than that of the lobe; i.e., $\Delta \rho \sim \rho_s$. 
%
%
%
%
%
Meanwhile, the longest Rayleigh-Taylor mode is taken as the local radius of curvature of the lobe. That is, for the ellipsoidal lobe geometry assumed in the RAiSE model, the radius of curvature at polar angle $\theta$ is given by,
\begin{equation}
\varrho(\theta) = \frac{R(\theta)}{A^2 \eta^4(\theta)},
\end{equation}
where $A$ is the late-time axis ratio of the lobe, and $\eta(\theta)$ is a geometric factor defined in Equation 14 of \citet{Turner+2023a}. 
The expression for the critical magnetic field strength becomes:
\begin{equation}
B_\text{crit} = \sqrt{\frac{\;\!\mu_0 \beta(r) p(r) \rho_s(r)}{2\pi A^2 \eta^4(\theta) \rho_{\rm x}(r)}} \leqslant \sqrt{\frac{\;\!\mu_0 A^2 \beta(r) p(r) \rho_s(r)}{2\pi \rho_{\rm x}(r)}},
\label{B crit}
\end{equation}
where we have applied the stated simplifications and used our earlier expression for the local gravitational acceleration (\cref{gravity}). The second inequality comes from noting that $\eta \rightarrow 1$ along the jet axis, and $\eta \rightarrow 1/A$ along the minor axis of the lobe.

The magnetic sheath will prevent (or greatly suppress) Rayleigh-Taylor instabilities if $B > B_\text{crit}$. The two expression for the magnetic field (\cref{B field,B crit}) yield the following condition for the magnetic suppression of the fluid instabilities:
\begin{equation}
\frac{\rho_s}{\rho_{\rm x}} < \frac{4\pi q}{A^2 \beta(r)(\Gamma_{\rm c} - 1)(q + 1)} .
\end{equation}
The shocked gas shell is at least as dense as the surrounding ambient medium but is not expected to exceed four times the ambient density based on the Rankine–Hugoniot shock jump conditions for a monotomic gas; i.e., $1 \leqslant \rho_s/\rho_{\rm x} \leqslant 4$. The inequality can therefore only be satisfied for strong, near equipartition fields, and towards the core of the cluster ambient medium when $\beta \ll 1$. Specifically, for $A=2.83$ (as assumed in this work) and $\rho_s/\rho_{\rm x} = 4$, we require (e.g.) an equipartition field and $\beta \leqslant 0.3$. Importantly, shorter wavelength modes will still be suppressed if the sheath magnetic field strength is somewhat close to the critical value, preventing the small-scale turbulent shredding that has the fastest growth rates.


The range of plausible values suggest that an effective magnetic surface tension may somewhat stabilise the interface between the lobe and shocked shell relative to the hydrodynamic simulation predictions (see \cref{sec:diffusion and fluid instabilities}), albeit, not to the extent of the contact discontinuity assumed in the analytical model. Remnant lobes in massive clusters with the shortest preceding active ages are most likely to be stabilised by a magnetic sheath as these objects remain in the flatter cluster core. The exact behaviour at this interface cannot be robustly addressed without a suite of magnetohydrodynamic simulations covering a range of magnetic field strengths, well beyond the scope of this work.

\subsection{Turbulent mixing from fluid instabilities}
\label{sec:diffusion and fluid instabilities}

The RAiSE dynamical model assumes a contact discontinuity between the lobe and shocked gas shell; this assumption may not remain valid in remnant lobes with only weak magnetic fields at the interface with the dense shocked gas shell, or at higher cluster-centric radii (\cref{sec:magnetic field surface tension}). The amplitude of oscillations at the interface due to fluid instabilities (Rayleigh-Taylor and Kelvin-Helmholtz) will increase exponentially without the presence of a damping magnetic field. Similarly, diffusion of particles can occur unabated without a formal boundary between the two media.

The analytic theory presented in this work is compared to a hydrodynamic simulation (without magnetic fields) to investigate the importance of fluid instabilities in non-magnetically-damped bubbles. The simulation in this work uses the publicly-available \textsc{pluto} code for astrophysical gas dynamics \citep{Mignone+2007}; details of the numerical implementation are the same as presented in \citet{Stewart+2025}. We assume each jet has a 10~degree half-opening angle, a Lorentz factor of $\gamma = 5$, and a jet power of $Q = 5\times10^{35}$~W (i.e., $Q_\text{tot} = 10^{36}$~W). The ambient medium is modelled with the default density and temperature profiles assumed in RAiSE for a $10^{14}$~M$_\odot$ cluster. The simulated jet/lobe has an active age of $t_\text{on} = 10$~Myrs; this location in parameter space is marked on \cref{fig:tcrit} by a star.

\begin{figure}
\centering
\includegraphics[width=\columnwidth,trim={15 7 35 40},clip]{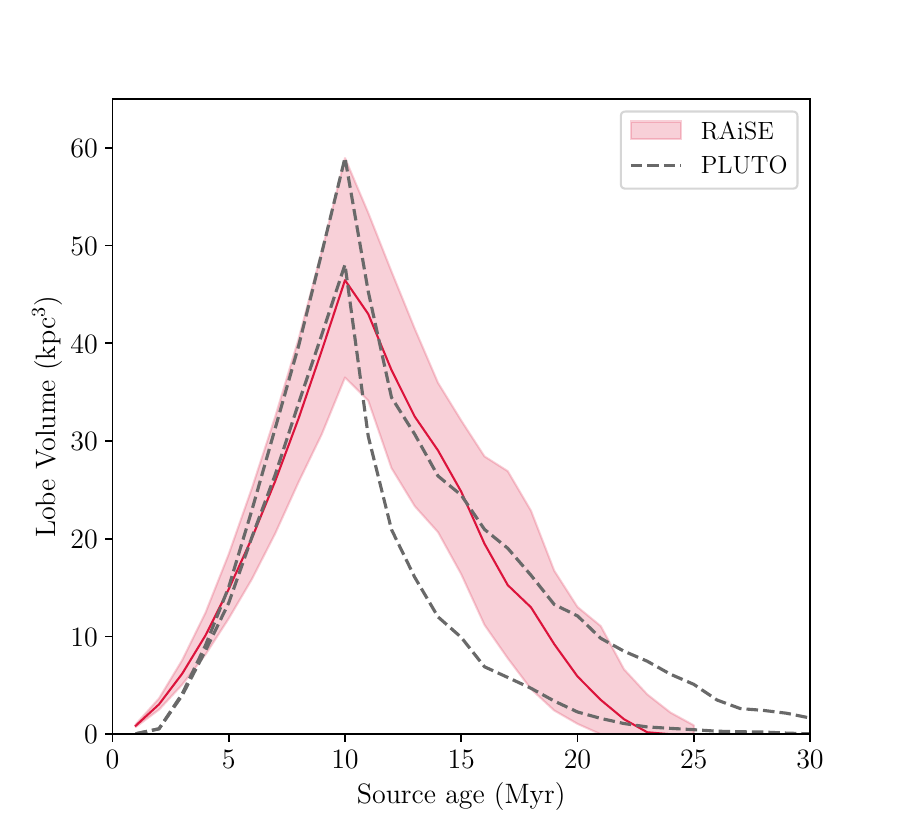}
\caption{Comparison of imploding remnant dynamics using RAiSE and the PLUTO hydrodynamic simulation package. The model is applied for a $10^{36}$~W jet power in a $10^{14}$~M$_\odot$ cluster with an active age of 10~Myrs. The shaded region shows the uncertainty in the volume for a 0.1~dex systematic error in the cluster mass. The two dashed lines correspond to two plausible mass tracer cuts in the hydrodynamic simulation.}
\label{fig:hydro}
\end{figure}

The RAiSE dynamical model accurately captures the lobe length evolution during the active phase, however, the volume of the lobe is somewhat sensitive to the mass tracer cut due the mixing of particles across the `contact discontinuity'. The lobe volumes are consistent in the active phase for a mass tracer cut of $\psi = 0.015$-$0.02$, as shown in \cref{fig:hydro}. The volumes produced with this mass tracer cuts in this range correspond to an axis ratio of approximately $A = 2.8$; this is consistent with expectations for an jet half-opening angle of 10~degrees \citep[e.g.,][]{Komissarov+1998, Yates+2022}, in addition to the assumed value in RAiSE. Lower mass tracer cuts yield a lobe volume that converges to that of a spherical lobe with the jet length predicted by RAiSE, suggesting a small fraction of jet plasma mixes beyond the `contact discontinuity' in all directions to a given galactocentric radius. 

The behaviour of our modified dynamical model (\cref{sec:buoyancy-driven expansion}) and the hydrodynamic simulation in the remnant phase is expected to differ as the analytical model only includes an approximate treatment of buoyancy, in addition to the contact discontinuity enforced between the lobe and shocked shell. Regardless, the volume of the lobe after the cessation of jet activity (at $10$~Myr) is consistent between our modified dynamical model and the hydrodynamic simulation (for the $\psi = 0.015$-$0.02$ mass tracer cut), as shown in \cref{fig:hydro}. 

Crucially, fluid instabilities will lead to a minor reduction in density in the region surrounding the lobe, and thus, the state of the ambient medium after `implosion' may differ from our expectation assuming the presence of a contact discontinuity; i.e., fluid instabilities will lead to a broad depression in the ambient gas density profile whereas a `clean' implosion yields the equilibrium ambient profile. Either way, the volume of plasma with a mass tracer sufficiently high for any appreciable synchrotron radiation quickly falls to near-zero, preventing the remnant from being detected at radio frequencies.

\section{Conclusion}
\label{sec:conclusion}

We have proposed a physical mechanism to explain the observed scarcity of lobed remnant AGNs in massive clusters \citep[e.g.,][]{Jurlin+2021, Singh+2021, Dutta+2023}. The key equations of the \textit{Radio AGN in Semi-analytic Environments} \citep[RAiSE;][]{Turner+2023a} dynamical model are investigated using asymptotic analysis to characterise the limiting behaviour of remnant lobes in regions of jet power--ambient medium parameter space. We find two important limiting cases: (1) lobe and shocked gas shell momentum driven expansion, $v_s \gg c_{\rm x}$ (here $v_s$ is the velocity of the shocked shell, and $c_{\rm x}$ is the sound speed of the ambient medium); and (2) ambient pressure confined expansion, $v_s \ll c_{\rm x}$. The former case, associated with powerful jets in the preceding active phase, leads to slow remnant growth of the form $R_s(t) \propto t^{2/(7 - \beta)}$ (\cref{solution 1}; here, the ambient gas density is $\rho \propto r^{-\beta}$). The latter case, associated with weaker jets, leads to a rapid `implosion' to zero radius on a timescale of order a few Myr (\cref{implosion timescale}). 

We derive an analytic relationship for a critical line of stability below which remnant lobes are unstable to implosion (\cref{critical line}):
\begin{equation*}
\begin{split}
&\left[\frac{Q_\text{on}}{10^{35.5}\,\rm W} \right]\!\;\! \left[\frac{\rho(a = 10\rm\,kpc)}{5\times10^{-24}\,\rm kg\, m^{-3}} \right]^{-1\!\;\!} \left[\frac{T}{10^7\rm\, K} \right]^{(\beta - 5)/2} \left[\frac{t_\text{on}}{10\rm\, Myr} \right]^{\!\;\beta - 2} \approx 1,
\end{split}
\end{equation*}
where $Q_\text{on}$ is the jet power in the preceding active phase (each jet; i.e., $Q_\text{tot} = 2Q_\text{on}$), $\rho$ is the density of the ambient medium at a galactocentric radius of $10$~kpc, $T$ is the temperature of the cluster, and $t_\text{on}$ is the duration of the active phase.

We calculate remnant implosion timescales, $t_\text{crit}$, across jet power--ambient medium ($Q_\text{tot}, M_\text{halo}$) parameter space for three preceding active ages, $t_\text{on}$, using the full RAiSE dynamical model (\cref{sec:results}). We discuss potential limitations of the dynamical model based predictions and compare our results to a hydrodynamic simulation in the region of the parameter space unstable to remnant implosion (\cref{sec:discussion}). The key findings from our analysis are as follows:
\begin{itemize}    
    \item Low-powered lobed AGNs and those in massive clusters are unstable to implosion in their remnant phase (\cref{sec:remnant implosion timescales}); such objects are visible for less one-tenth the time of stable lobes ({\cref{fig:visible}}).
    \item Remnant lobed AGNs in massive clusters ($M_\text{halo} \sim 10^{14.5}$~M$_\odot$) will be under-counted by a factor of at least five compared to those in poorer groups ($M_\text{halo} \sim 10^{12}$~M$_\odot$; \cref{fig:hist} and \cref{sec:remnant visibility}).
    \item RAiSE model predictions are consistent with an equivalent hydrodynamic simulation for an appropriate mass tracer cut reflecting the turbulent mixing in the active phase (\cref{fig:hydro} and \cref{sec:diffusion and fluid instabilities}).
    \item Strong magnetic fields in the lobe-sheath {may} prevent turbulent mixing
    for remnant lobes in cluster cores, with fast-growing short-wavelength shredding suppressed for weaker fields (\cref{sec:magnetic field surface tension}); this leads to a `clean' implosion (\cref{sec:diffusion and fluid instabilities}).
    \item Remnant lobes without a stabilising magnetic field will have fluid instabilities turbulently mix the lobe plasma and shocked gas shell; this leads to a broad depression in the ambient gas density profile (\cref{sec:diffusion and fluid instabilities}).
    \item Buoyant bubbles will not rise out of the cluster potential if the lobe-sheath magnetic field permits a `clean' implosion ({\cref{fig:buoyancy} and \cref{sec:buoyancy-driven expansion}}); e.g., high sheath magnetic field and lobe in cluster core ({\cref{sec:magnetic field surface tension}}).
\end{itemize}

The missing population of low-powered remnant lobes predicted in this work, especially in massive clusters, is of critical importance for studies of the energetics of AGN feedback. The number of low-powered remnant lobes (e.g., $Q_\text{tot} = 10^{35}$~W) is under-counted by a factor at least ten compared to higher jet powers (e.g., $Q_\text{tot} = 10^{38}$~W) for preceding active ages of 1 and 10~Myr (see \cref{fig:visible}); this is in addition to the bias against massive clusters. These objects are only a few to tens of kpc in total size, occupying a region of size--luminosity parameter space with a known deficit of extended radio AGNs \citep[e.g.,][their Figures 6 and 8]{Hardcastle+2019}. The lack of these remnant AGNs in observed populations will consequently lead to a severe underestimate of the number of low-powered jet outbursts, and skew observed correlations between jet power and cluster environment.
Moreover, AGN feedback is most effective towards the rapidly cooling cluster cores. The presence of a potentially large population of `missing' AGNs at these small galactocentric radii is therefore especially important for reconciling predictions of semi-analytic galaxy evolution models with observations.

\section*{Acknowledgements}

{We thank the anonymous referee for a constructive and prompt report which helped improve the manuscript.}





\bibliographystyle{elsarticle-harv}
\bibliography{implodingremnants} 




\appendix

\section{Relation to large-scale host cluster parameters}
\label{sec:relation to large-scale host cluster parameters}

The condition for the first asymptotic solution in \cref{critical line} is not represented in terms of the most helpful parameters for comparing with observed radio sources. We express the ambient gas density and temperature profiles in terms of the cluster mass as follows:
\begin{equation}
\begin{split}
M_\text{halo} = \int_0^{r_\text{halo}}\;\!\!  \frac{\rho}{f_\text{gas}}\;\!\! \left(\frac{r}{a}\right)^{-\beta'}\;\!\! \text{d}V \quad = \frac{4\pi a^{\beta'} \,\;\!\!\rho}{(3 - \beta')f_\text{gas}} r_\text{halo}^{3 - \beta'},
\end{split}
\label{mass 1}
\end{equation}
where $r_\text{halo}$ is the virial radius (e.g., radius for an overdensity of 200), $f_\text{gas}$ is the average baryon fraction within the cluster, and $\beta'$ is an appropriate power law exponent to approximate the mass integral of the full ambient density profile over the specified radial range.
Then, using the redshift-dependent relationship between the cluster mass and virial radius \citep[e.g.,][their Equation 2]{Croton+2006}, we obtain an expression for the core density as follows:
\begin{equation}
\begin{split}
\rho = \frac{(3 - \beta')f_\text{gas}}{4\pi a^{\beta'}}\left[\frac{100H^2(z)}{G} \right]^{(3 - \beta')/3} M_\text{halo}^{\beta'/3} ,
\end{split}
\label{mass 2}
\end{equation}
where $G$ is the gravitational constant, and $H(z)$ is the Hubble constant at redshift $z$.

The temperature of a cluster in hydrostatic equilibrium, outside the core \cite[e.g.,][]{Sullivan+2025}, is directly related to the total mass as
\begin{equation}
\begin{split}
T = T_0 \left[\frac{M_\text{halo}}{M_0} \right]^{\alpha} ,
\end{split}
\label{temperature}
\end{equation}
for some chosen critical temperature $T_0$, and a critical cluster mass $M_0$ informed by X-ray observations \citep[e.g.,][]{Vikhlinin+2006}. The power-law slope is known to be $\alpha = 2/3$ for massive clusters in hydrostatic equilibrium, however, a weaker relationship is apparent poorer groups (e.g., \citealt{Sun+2009}; and discussion in \citealt{Turner+2015}).

The bound for the first asymptotic solution (\cref{critical line}) can now be expressed entirely in terms of the jet power, cluster mass, and source age as follows:
\begin{equation}
\begin{split}
&Q_\text{on} M_\text{halo}^{f(\alpha,\,\beta,\,\beta')} t_\text{on}^{\!\;\beta - 2} \gg \frac{9(3 - \beta')(11 - \beta)f_\text{gas}}{4 a^{\beta'\! - \beta} (5 - \beta)^3} \\ &\qquad\qquad \times \bigg[\frac{(5 - \beta)^2k_{\rm B} T_0}{9\bar{m} M_0^\alpha} \bigg]^{(5 - \beta)/2} \left[\frac{100H^2(z)}{G} \right]^{(3 - \beta')/3}\! ,
\end{split}
\label{critical line 2}
\end{equation}
where $f(\alpha,\beta,\beta') = [3\alpha(\beta - 5) - 2\beta']/6$.

We simplify this relationship, for clarity of our argument, by assuming the gas fraction is $f_\text{gas} = 0.15$ (i.e., approaches the cosmic baryon fraction), the scale-radius of the density profiles is $a = 10$\;kpc, the critical temperature and mass are $T_0 = 10^{7}$\;K and $M_0 = 10^{13.63}$\;$\text{M}_\odot$ \citep[mean halo mass at $10^7$\;K in RAiSE model; cf.][]{Vikhlinin+2006}, respectively, and that the redshift-dependent Hubble constant is $H(z) \approx H_0 = 70$\;km\;s$^{-1}$\;Mpc$^{-1}$ in the local universe (i.e., $z \ll 1$).
The bound for the the first asymptotic solution thus becomes:
\begin{equation}
\begin{split}
&\left[\frac{Q_\text{on}}{10^{35.5}\,\rm W} \right]\!\;\! \left[\frac{M_\text{halo}}{10^{13.63}\rm\, M_\odot} \right]^{f(\alpha,\,\beta,\,\beta')} \left[\frac{t_\text{on}}{10\rm\, Myr} \right]^{\!\;\beta - 2} \\
&\qquad\quad \gg 0.0139^{1.526 - \beta'\!} (3 - \beta') (11 - \beta) \big[\!\; 0.1264 (5 - \beta) \big]^{2 - \beta},
\end{split}
\label{critical line 3}
\end{equation}
where the right-hand side of this expression is moderately sensitive to the average power-law density profile of the cluster, however, this power-law slope is observed to take a relatively restricted set of values with mean and standard deviation $\beta' = 0.76\pm 0.09$ \citep[e.g.,][their Table 2]{Vikhlinin+2006}.

\label{lastpage}
\end{document}